\documentclass[acmsmall,screen]{acmart}

\usepackage{listings}
\usepackage[most]{tcolorbox}
\usepackage{subcaption}
\usepackage{xcolor}
\usepackage[normalem]{ulem} 
\usepackage{multirow}
\usepackage{multicol}
\usepackage{booktabs}

\newcommand{\ea}{\textit{et al.}}

\newcommand{\find}[1]{
\begin{tcolorbox}[leftrule=1mm,toprule=0mm,bottomrule=0mm,left=1pt,right=2pt,top=2pt,bottom=2pt]
\em #1
\end{tcolorbox}
}

\newfloat{listing}{htbp}{lol}
\floatname{listing}{Listing} 

\definecolor{issuebg}{HTML}{fff5f0}
\definecolor{issueframe}{HTML}{fb6a4a}
\definecolor{fixbg}{HTML}{F0FDF4}
\definecolor{fixframe}{HTML}{15803D}
\definecolor{codekw}{HTML}{1D4ED8}
\definecolor{codecom}{HTML}{15803D}
\definecolor{codestr}{HTML}{B42318}
\definecolor{codetext}{HTML}{111827}
\definecolor{buglinecolor}{HTML}{cb181d}
\newcommand{\bugline}[1]{%
  \mbox{\color{buglinecolor}\uwave{\color{codetext}#1}}
}

\lstdefinestyle{compactpython}{
  language=Python,
  basicstyle=\ttfamily\scriptsize\color{codetext},
  keywordstyle=\color{codekw}\bfseries,
  commentstyle=\color{codecom}\itshape,
  stringstyle=\color{codestr},
  showstringspaces=false,
  showspaces=false,
  showtabs=false,
  keepspaces=true,
  breaklines=true,
  breakatwhitespace=false,
  columns=fullflexible,
  tabsize=4
}

\newtcblisting{issuecode}[2][]{%
  enhanced,
  listing only,
  colback=issuebg,
  colframe=issueframe,
  boxrule=0.55pt,
  arc=0.9mm,
  left=0.35mm,
  right=0.35mm,
  top=-0.15mm,
  bottom=-0.15mm,
  title={#2},
  fonttitle=\bfseries\scriptsize,
  coltitle=black,
  attach boxed title to top left={xshift=1mm,yshift=-0.8mm},
  boxed title style={
    colback=issuebg,
    colframe=issueframe,
    boxrule=0.55pt,
    arc=0.9mm,
    left=0.6mm,
    right=0.6mm,
    top=0.3mm,
    bottom=0.3mm
  },
  listing options={
    style=compactpython,
    escapechar=|,
    aboveskip=0pt,
    belowskip=0pt
  },
  #1
}

\newtcblisting{fixcode}[2][]{%
  enhanced,
  listing only,
  colback=fixbg,
  colframe=fixframe,
  boxrule=0.55pt,
  arc=0.9mm,
  left=0.35mm,
  right=0.35mm,
  top=-0.15mm,
  bottom=-0.15mm,
  title={#2},
  fonttitle=\bfseries\scriptsize,
  coltitle=black,
  attach boxed title to top left={xshift=1mm,yshift=-0.8mm},
  boxed title style={
    colback=fixbg,
    colframe=fixframe,
    boxrule=0.55pt,
    arc=0.9mm,
    left=0.6mm,
    right=0.6mm,
    top=0.3mm,
    bottom=0.3mm
  },
  listing options={
    style=compactpython,
    aboveskip=0pt,
    belowskip=0pt
  },
  #1
}

\newcommand{\smallsection}[1]{\noindent {\bf \underline{#1}}.\hspace{1mm}}

\begin{document}

\title{Debt Behind the AI Boom: A Large-Scale Empirical Study of AI-Generated Code in the Wild}

\author{Yue Liu}
\affiliation{%
  \institution{Adelaide University}
  \city{Adelaide}
  \state{South Australia}
  \country{Australia}
}
\email{knox.liu@adelaide.edu.au}

\author{Ratnadira Widyasari}
\affiliation{%
  \institution{University of Victoria}
  \city{Victoria}
  \country{Canada}
}
\email{ratnadiraw@uvic.ca}

\author{Yanjie Zhao}
\affiliation{%
  \institution{Huazhong University of Science and Technology}
  \city{Wuhan}
  \country{China}
}
\email{yanjie_zhao@hust.edu.cn}

\author{Ivana Clairine Irsan}
\affiliation{%
  \institution{Singapore Management University}
  \city{Singapore}
  \country{Singapore}
}
\email{ivanairsan@smu.edu.sg}

\author{Junkai Chen}
\affiliation{%
  \institution{Singapore Management University}
  \city{Singapore}
  \country{Singapore}
}
\email{junkaichen@smu.edu.sg}

\author{David Lo}
\affiliation{%
  \institution{Singapore Management University}
  \city{Singapore}
  \country{Singapore}
}
\email{davidlo@smu.edu.sg}

\renewcommand{\shortauthors}{Liu et al.}

\begin{abstract}
AI coding assistants are now widely used in software development. 
Software developers increasingly integrate AI-generated code into their codebases to improve productivity.
Prior studies have shown that AI-generated code may contain code quality issues under controlled settings.
However, we still know little about the real-world impact of AI-generated code on software quality and maintenance after it is introduced into production repositories.
In other words, it remains unclear whether such issues are quickly fixed or persist and accumulate over time as technical debt.
In this paper, we conduct a large-scale empirical study on the code quality issues introduced by AI coding assistants in the wild.
To achieve that, we built a dataset of 302.6k verified AI-authored commits from 6,299 GitHub repositories, covering five widely used AI coding assistants.
For each commit, we run static analysis before and after the change to precisely attribute which code smells, correctness issues, and security issues the AI introduced.
We then track each introduced issue from the introducing commit to the latest repository revision to study its lifecycle.
Our results show that we identified 484,366 distinct issues, and that code smells are by far the most common type, accounting for 89.3\% of all issues.
Between 7.5\% and 14.8\% of each assistant's commits introduce at least one issue. 
More importantly, 22.7\% of tracked AI-introduced issues still survive at the latest version of the repository.
These findings show that AI-generated code can introduce long-term maintenance costs into real software projects and highlight the need for stronger quality assurance in AI-assisted development.

\end{abstract}

\maketitle


\section{Introduction}
Through AI coding assistants (e.g., Cursor, Claude Code), software developers can now describe what they want in natural language and get working code back in seconds.
They significantly improve development productivity.
Thus, AI is becoming standard equipment for modern software developers~\cite{github2024octoverse}.
According to the 2025 Stack Overflow Developer Survey, 84\% of professional developers are using or are planning to use AI coding tools within their development processes~\cite{stackoverflow2025survey}.
AI-generated code is also widely used in real-world software projects.
For example, both Google and Microsoft disclosed in 2025 that AI now writes over 20\% of their new code~\cite{novet2025microsoft, robison2024google}.
Similarly, GitHub reported that more than 1.1 million public repositories used AI coding tools between 2024 and 2025 ~\cite{github2024octoverse}.
Overall, AI-generated code has graduated from experiment to production reality, and it is everywhere.

Although AI coding assistants have proven effective at generating functional programs, many previous research studies have revealed a range of quality concerns in AI-generated code.
Recent studies have shown that AI-generated code suffers from functional bugs, runtime errors, and systemic maintainability issues~\cite{liu2024refining, siddiq2024quality}.
Also, the code produced by AI coding tools poses security risks~\cite{pearce2025asleep, perry2023users, liu2025ai}.
Pearce~\ea~\cite{pearce2025asleep} found that about 40\% of AI-generated code in security-sensitive contexts contains critical vulnerabilities.
However, recent research~\cite{perry2023users, sabouri2025trust} has found that developers tend to place excessive trust in the quality of AI-generated code, blindly accepting the code without proper validation.
As a result, these unverified code snippets are merged into production codebases.
Over time, this can cause a considerable accumulation of \textbf{technical debt}, which can be costly and time-consuming to address~\cite{harding2024coding, moreschini2026evolution}.

Recognizing these risks, recent empirical studies have started to investigate code quality issues of AI-generated code in real-world repositories. 
These studies cover a range of practical concerns, such as security weaknesses~\cite{fu2025security, wang2025ai}, project-level development velocity~\cite{he2025does}, pull request acceptance rates~\cite{watanabe2025use}, and code redundancy~\cite{huang2026more}.
For example, He~\ea~\cite{he2025does} found that Cursor adoption in 807 GitHub repositories led to a transient velocity boost but persistent increases in code complexity. 
However, existing studies still have several limitations.
First, most studies~\cite{watanabe2025use, he2025does} focus on a single tool or a narrow set of tools, which limits the generalizability of their findings.
Second, many studies~\cite{he2025does, watanabe2025use, huang2026more} do not identify AI-generated code at the commit level.
Instead, they rely on project-level adoption signals, such as the presence of AI tool configuration files in a repository (e.g., \textit{.cursorrules} or \textit{.claude/}).
These signals suggest possible AI use, but do not directly show which commits or files were generated by AI.
Third, they mostly evaluate code at a single point in time, failing to capture the long-term lifecycle of the code~\cite{fu2025security, liu2024refining, huang2026more}.
Finally, in real-world repositories, AI-assisted code and human-written code are often mixed together, and AI usage may leave no reliable trace.
Thus, it remains unknown how AI-generated code actually ages in production. 
We still do not know whether these code quality issues persist, get refactored, or silently accumulate over time.

To bridge this gap, we conduct a large-scale empirical study to investigate AI-introduced code quality issues in the wild.
Our approach consists of three steps.
First, we build a large dataset of verified AI-authored commits from five popular coding assistants (i.e., GitHub Copilot, Claude, Cursor, Gemini, and Devin) across over 6,000 GitHub repositories.
We use explicit Git metadata to identify commits generated by AI coding tools across thousands of GitHub repositories.
This design is analogous in spirit to research on self-admitted technical debt (SATD), which studies the subset of technical debt that developers explicitly document rather than the full universe of debt~\cite{potdar2014exploratory, da2017using}.
Similarly, we focus on the subset of AI-assisted contributions whose AI involvement is explicitly visible in Git metadata.
Although this does not cover all AI-assisted changes, it provides a more reliable basis for attribution and large-scale empirical analysis.
Second, we perform a commit-level quality analysis.
For each AI-authored commit, we run static analysis tools on the source code immediately before and after the change.
This allows us to precisely identify which code smells, correctness issues, and security issues the AI introduced or fixed.
Third, we conduct a debt lifecycle analysis.
We track each introduced issue to the latest repository revision (\texttt{HEAD}) to determine whether it still survives or has been resolved.
Through this design, our study provides the first comprehensive view of how AI-generated code ages in real-world software.

\noindent\textbf{Contribution.} To the best of our knowledge, this paper is the first to:
\begin{itemize}
    \item Conduct a large-scale empirical study of AI-introduced code quality issues across five major AI coding assistants (i.e., GitHub Copilot, Claude, Cursor, Gemini, and Devin) and over 6,000 real-world GitHub repositories.
    \item Perform commit-level differential analysis to precisely attribute code smells, correctness issues, and security issues to individual AI-authored commits.
    \item Track AI-introduced issues from the introducing commit to the latest repository revision, revealing whether it persists or gets resolved over time.
\end{itemize}

\textbf{\underline{Open Science.}} To support the open science initiative, we publish the study dataset and a replication package, which is publicly available in GitHub.\footnote{https://github.com/yueyueL/tech-debt-ai-coding}

\textbf{\underline{Paper Organization.}}
Section~\ref{sec:Background} presents the background and motivation.
Section~\ref{sec:methodology} describes our approach.
Section~\ref{sec:experiment} presents the experimental setup.
Section~\ref{sec:results} presents the results.
Section~\ref{sec:discussion} discusses the findings and their implications.
Section~\ref{sec:relatedwork} presents the related work.
Section~\ref{sec:threats} discloses the threats to validity.
Section~\ref{sec:conclusion} draws the conclusions.
 
\section{Background}
\label{sec:Background}

\subsection{AI Coding Assistants}
\begin{figure}[t]
    \centering
    \includegraphics[width=0.6\linewidth]{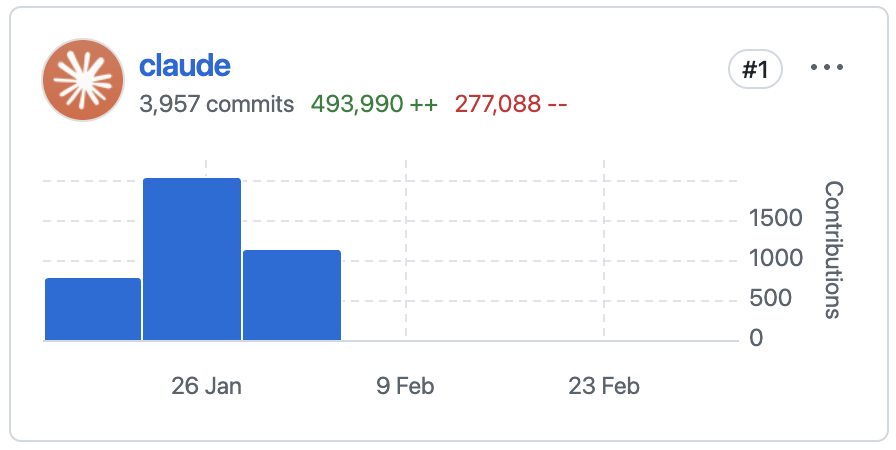} 
    \caption{Contributor statistics for the Anthropic \texttt{claudes-c-compiler} repository.}
    \label{fig:claude_commits}
\end{figure}

Modern AI coding assistants (e.g., Cursor, Claude Code, GitHub Copilot) are now deeply embedded in software development workflows.
They help software developers write, modify, explain, test, or debug code using natural-language instructions and code context.
Advanced agentic tools can now process whole functions, files, or even repositories to autonomously create pull requests with minimal human intervention.
Driven by these improved capabilities, AI-generated code is entering production codebases at unprecedented speed and scale. According to GitHub, over 1.1 million public repositories adopted AI coding tools between 2024 and 2025~\cite{github2024octoverse}.
As shown in Figure~\ref{fig:claude_commits}, in Anthropic's \texttt{claudes-c-compiler} repository~\cite{anthropic2026claudes-c-compiler}, Claude appears as the top contributor, with 3,957 commits and nearly 500K lines of code added within just a few weeks.
At this volume and speed, it is unlikely that all AI-generated code receives a thorough human review.
This makes it increasingly important to understand the long-term implications of AI-generated code on software quality and maintenance.

In controlled (lab) settings~\cite{pearce2025asleep,liu2024refining}, the provenance of code is usually clear since researchers can generate the code directly. 
However, it is different in real-world codebases.
AI-generated code and human-written code are often interleaved during development, and AI usage is not always explicitly recorded in the repository history.
This makes it harder to attribute code changes reliably and to observe the long-term impact of AI-generated code after it is merged into production codebases.



\subsection{Technical Debt}
Technical debt refers to design or implementation choices that prioritize short-term speed over long-term quality~\cite{cunningham1992wycash}. 
These shortcuts may help in the short term,  but they increase the future cost of maintaining and evolving the software~\cite{avgeriou2016managing, li2015systematic}.
Its costs can accumulate over time if not addressed.
This concern becomes more important as AI coding assistants are widely adopted.
AI coding assistants help developers work faster and produce more code.
Previous studies have shown that AI-generated code contains code smells, correctness issues, and security vulnerabilities~\cite{liu2024refining, pearce2025asleep, fu2025security}.
In this study, we focus on these three categories as code-level technical debt: code smells that reduce maintainability, correctness issues that affect program behavior, and security issues that may expose systems to risk.
When such issues are accepted into production repositories, they can accumulate as technical debt in the codebase.

\begin{figure}[t]
\centering

\begin{issuecode}{\texttt{hysteria2~\cite{hysteria2repo}}, \texttt{hysteria2.py:L77}, \texttt{e277daf}}
# issue: shell=True introduces command-injection risk
subprocess.run(
    "systemctl stop hysteria-iptables.service 2>/dev/null",
    |\bugline{shell=True}|
)
\end{issuecode}

\vspace{-0.9em}
\noindent\hspace*{-20.0em}{\large$\Downarrow$}
\vspace{-0.8em}

\begin{fixcode}{\texttt{hysteria2~\cite{hysteria2repo}}, \texttt{hysteria2.py:L77} \texttt{d9e392d}}
#fix: ``Improve code security by removing shell=True from subprocess calls''
subprocess.run(
    ["systemctl", "disable", "hysteria-iptables.service"],
    stderr=subprocess.DEVNULL
)
\end{fixcode}

\caption{Command injection risk introduced by GitHub Copilot in \texttt{hysteria2} (1.7K stars)~\cite{hysteria2repo} and later removed in a Copilot-assisted fix commit~\cite{hysteria2commit_fix}.}
\label{fig:hysteria2_example}
\end{figure}

\begin{figure}[t]
\centering

\begin{issuecode}{\texttt{librealsense~\cite{librealsenserepo}}, \texttt{test-fps-performance.py:L168}, \texttt{5535b8a}}
# issue: undefined constant used before definition
|\bugline{time.sleep(DEVICE\_INIT\_SLEEP\_SEC)}|
\end{issuecode}

\vspace{-0.9em}
\noindent\hspace*{-20.0em}{\large$\Downarrow$}
\vspace{-0.8em}

\begin{fixcode}{\texttt{librealsense~\cite{librealsenserepo}}, \texttt{test-fps-performance.py:L37}, \texttt{14026c8}}
# fix: ``Add missing constants & fixed 6fps bug for live/FPS weekly CI''
DEVICE_INIT_SLEEP_SEC = 3
time.sleep(DEVICE_INIT_SLEEP_SEC)
\end{fixcode}

\caption{Undefined variables introduced by GitHub Copilot in \texttt{librealsense} (8.6K stars)~\cite{librealsenserepo}, causing a runtime error. A human maintainer committed a fix three weeks later~\cite{librealsensecommit_fix}.}
\label{fig:librealsense_example}
\end{figure}

\subsection{Motivation}
\label{sec:motivation}
The technical debt introduced by AI coding assistants is not just a theoretical risk.
Such issues are becoming increasingly common in real-world codebases.
Figure~\ref{fig:hysteria2_example} and~\ref{fig:librealsense_example} show two examples.
In Figure~\ref{fig:hysteria2_example}, a GitHub Copilot-authored commit introduced a \texttt{shell=True} subprocess call in \texttt{hysteria2.py}~\cite{hysteria2commit_issue}. 
This pattern increases security risk by allowing command injection if user input is involved.
A human developer later fixed it, noting in the commit message: \textit{``Improve code security by removing shell=True from subprocess calls''}~\cite{hysteria2commit_fix}.
Figure~\ref{fig:librealsense_example} shows a bug from Intel's \texttt{librealsense}~\cite{librealsenserepo}.
In another case, GitHub Copilot replaced a literal value with a named constant, but never defined the constant~\cite{librealsensecommit_issue}.
The buggy code remained in the repository for over three weeks before the maintainer committed a fix adding the missing definition~\cite{librealsensecommit_fix}.

These two examples highlight the motivation of our study.
AI coding assistants can generate functional code, but they may introduce quality issues into production codebases.
Developers also tend to over-trust and accept AI suggestions without thorough review~\cite{perry2023users, sabouri2025trust}.
These issues may be fixed later, or they may persist for a long time, or even create long-term maintenance challenges.
Recent studies~\cite{nguyen2022empirical, liu2024refining, fu2025security, siddiq2024quality, wang2025ai, he2025does, watanabe2025use} have examined the AI-generated code, but they have several limitations.
Most prior studies focus on a single tool, a small set of tasks, or controlled settings.
For example, Watanabe~\ea~\cite{watanabe2025use} measured the initial acceptance rate of AI-authored code in a single repository.
We still know little about the long-term implications of AI-generated code on software quality and maintenance.
Thus, we study the code quality issues introduced by various AI coding assistants in the wild: how often it is introduced, what kinds of issues appear, and whether those issues still remain in the codebase over time.

\section{Approach}
\label{sec:methodology}

\begin{figure*}[t]
    \centering    \includegraphics[width=1\textwidth]{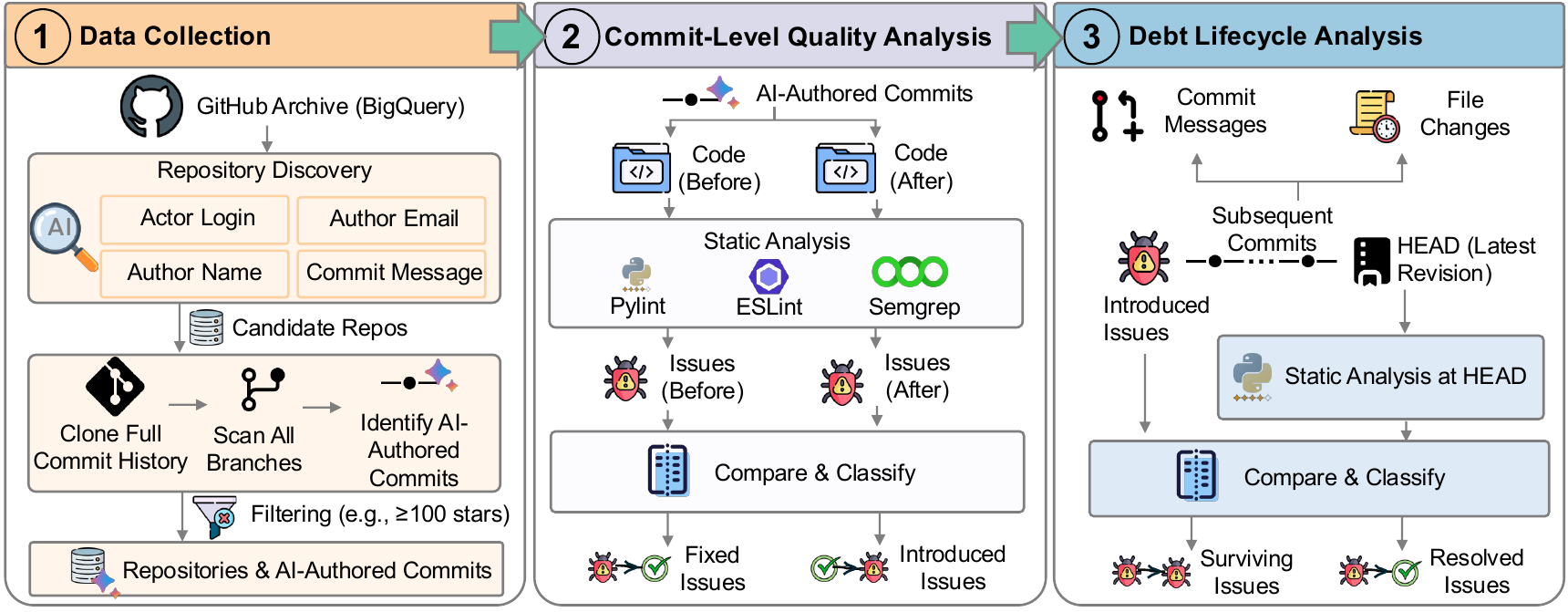}
    \caption{Overview of our approach.}
    \label{fig:pipeline}
\end{figure*}

Figure~\ref{fig:pipeline} provides an overview of our approach.
We first collect AI-authored commits from GitHub repositories at scale (Section~\ref{sec:data_collection}).
We then analyze each AI-authored commit at the code level to determine which quality issues it introduced or fixed (Section~\ref{sec:quality_analysis}).
Finally, we track the lifecycle of both the issues and the code itself to determine whether AI-introduced debt persists or gets resolved over time (Section~\ref{sec:lifecycle}).

\subsection{Data Collection}
\label{sec:data_collection}
This step aims to identify candidate GitHub repositories that contain AI-authored commits.

\smallsection{Repository Discovery}
We use the GitHub Archive dataset~\cite{githubarchive}, which records public GitHub events (e.g., \texttt{PushEvent}), to identify repositories with potential AI-authored code.
The dataset is publicly available through Google BigQuery~\cite{bigquerygharchive}, which allows large-scale querying over historical events.
We scan \texttt{PushEvent} records from January 2024 to March 2026, focusing on repositories with recent development activity.
From each event, we extract four metadata fields: actor login, author name, author email, and commit message.
We then match these fields against our curated AI-attribution rules (described in the next section) to identify potential AI-authored commits.
Only repositories with at least one matching event are retained.
To improve coverage of active and popular repositories, we also query the GitHub REST API~\cite{githubrestapi} for top-starred repositories and apply the same attribution rules through full-history repository scanning.
This step helps us ensure that repositories with substantial AI-authored activity are not missed during the discovery stage.


\smallsection{AI Attribution Rules}
We build attribution rules for widely adopted AI coding tools (e.g., Cursor, GitHub Copilot, Claude Code) identified in the 2025 Stack Overflow Developer Survey~\cite{stackoverflow2025survey}.
We identify AI-authored commits using explicit signals in Git metadata.
Our approach covers AI-authored commits only when the use of an AI coding tool leaves explicit traces in Git metadata.
The rules are based on four sources of evidence:
(1)~~actor logins (e.g., \nolinkurl{copilot-swe-agent[bot]}),
(2)~author emails (e.g., \nolinkurl{noreply@anthropic.com}),
(3)~author names (e.g., \texttt{Cursor Agent}),
and (4)~\texttt{Co-authored-by} trailers in commit messages. 
These rules rely on explicit machine-readable signals in Git metadata.
To finalize the rule list, two authors manually reviewed candidate patterns and verified that they provided reliable evidence of AI coding tool involvement.
In total, 29 AI coding tools left identifiable traces in the repositories we collected.
The full list of tools and detection rules is included in our replication package.

\smallsection{Full-History Commit Scanning}
The discovery stage captures only push-event metadata.
However, it provides only partial evidence about AI-authored activity in a repository.
To obtain a more complete set of AI-authored commits, we perform a bare clone of each candidate repository.
We then scan the full commit history across all branches and apply the same attribution rules to every commit.
For each commit, we extract the SHA, author and committer metadata, timestamp, and full commit message.
This step allows us to identify AI-authored commits that are not directly visible during repository discovery (e.g., commits on non-default branches, commits outside the observation window).

\smallsection{Filtering}
To focus on established open-source projects, we filter out repositories that do not meet our study criteria.
We keep only repositories with at least 100 GitHub stars.
We also require at least one confirmed AI-authored commit.
Our downstream analysis is restricted to production Python, JavaScript, and TypeScript source files, since these are among the most widely used programming languages~\cite{stackoverflow2025survey} and are well supported by static analysis tools.
We therefore exclude repositories that do not contain any source files in these languages.
In total, the discovery stage identified \textit{587,118} candidate repositories.
After applying the star threshold, \textit{12,770} repositories remained.
After full-history scanning and language filtering, we obtained \textit{6,699} repositories with confirmed AI-authored commits.

\subsection{Commit-Level Quality Analysis}
\label{sec:quality_analysis}
To understand the impact of AI coding tools, we avoid evaluating a repository snapshot at a single point in time. 
For each AI-authored commit $c$, we analyze two versions of the source code: the version at $c$'s parent revision (\textit{before} the commit is applied) and the version at $c$ itself (\textit{after} the commit is applied).
Comparing these two versions allows us to determine which quality issues the commit introduced or fixed.
Our analysis focuses on source files written in Python, JavaScript, and TypeScript.
We exclude files that are unlikely to reflect production code quality (e.g., tests, documentation, configuration files, auto-built artifacts, and vendored dependencies).
Files are classified based on their paths and naming patterns (e.g., files under \texttt{test/} or \texttt{\_\_tests\_\_/} directories, or matching \texttt{*\_test.py}).
The full classification rules are included in our replication package.

\smallsection{Static Analysis}
For each AI-authored commit $c$ that modifies a source file $f$, we check out two versions of $f$ (i.e., the version at $c$'s parent commit (\textit{before}) and the version after applying $c$ (\textit{after})).
We run the same static analysis toolchain on both versions to identify potential code issues.
We use ESLint (for JavaScript and TypeScript)~\cite{eslint} and Pylint (for Python)~\cite{pylint} to detect code smells and correctness issues.
For security-related issues, we use Semgrep~\cite{semgrep}, which provides a unified framework for multi-language static analysis.
For each detected issue, we record its rule identifier, line number, detector, and message.
Figure~\ref{fig:issue-record-example} shows one such record, where ESLint flags a duplicate object key in a Claude-authored commit~\cite{superagent_commit_46695d1}.
This produces two issue sets $I$: the issue set before the commit, denoted by $I_f^{-}$, and the issue set after the commit, denoted by $I_f^{+}$.

\begin{figure}[t]
\begin{tcolorbox}[
  colback=gray!5,
  colframe=gray!60,
  boxrule=0.4pt,
  arc=2pt,
  left=4pt, right=4pt, top=3pt, bottom=3pt,
  fontupper=\footnotesize\ttfamily
]
Repository: superagent-ai/superagent \\
Commit: 46695d1 (author: Claude) \\
File: node/src/server.js \\
Detector: ESLint \\
Rule: no-dupe-keys \\
Line: 86 \\
Message: Duplicate key `lazyConnect'.
\end{tcolorbox}
\caption{Example of a recorded issue detected by ESLint in a Claude-authored commit~\cite{superagent_commit_46695d1}.}
\label{fig:issue-record-example}
\end{figure}

\smallsection{Differential Attribution}
To find out which issues commit $c$ introduced or fixed, we compare $I_f^{-}$ and $I_f^{+}$.
We also use \texttt{git diff} to extract the set of changed lines, denoted by $\Delta_f$.
However, not all differences between $I_f^{-}$ and $I_f^{+}$ (i.e., issues in $I_f^{+} \setminus I_f^{-}$ or $I_f^{-} \setminus I_f^{+}$) represent real changes.
When a commit inserts or deletes lines, it can cause existing issues to shift line numbers.
To address this, we first match issues across the two sets.
An issue $i$ is considered matched if the same rule and message appear in both $I_f^{-}$ and $I_f^{+}$ at the same or nearby line number.
After matching, the remaining issues are classified as follows.
An unmatched issue $i \in I_f^{+}$ is classified as \textit{introduced} only if its line falls within $\Delta_f$.
This means the issue exists only after the commit, on a line that the commit actually changed.
An unmatched issue $i \in I_f^{-}$ is classified as \textit{fixed}.
This means the issue existed before the commit but is no longer present afterward.

\subsection{Debt Lifecycle Analysis}
\label{sec:lifecycle}
Detecting technical debt at the time of introduction is only half the picture.
An issue that is quickly resolved has a very different cost than one that lingers for months.
We therefore track whether AI-introduced issues persist or get resolved over time.

\smallsection{Issue Survival}
For each issue introduced by an AI-authored commit, we check whether it still exists at the repository's latest revision (i.e., \texttt{HEAD}).
If the file has been renamed, we follow its history using \texttt{git log --follow}.
We then run static analysis on the corresponding file at \texttt{HEAD}.
Next, we look for the same issue in the analysis results.
We do not rely on the line number alone, since the location of the issue may move as the file changes.
Instead, we match issues using their rule identifier together with a small amount of surrounding code context.
If a match is found, the issue is classified as \textit{surviving}.
Otherwise, it is classified as \textit{not surviving}.
In other words, an introduced issue is counted as surviving only if the same issue is still present at \texttt{HEAD}. 
If the original issue disappears and a different issue appears later, the original issue is treated as \textit{not surviving}.

At the same time, we also record whether files touched by AI-authored commits are modified again before \texttt{HEAD}.
We trace the subsequent commit history of each affected file to understand how actively it is maintained after the AI-authored change.
This additional context helps us interpret the survival results and understand the maintenance patterns around AI-introduced debt.

\section{Experimental Setup}
\label{sec:experiment}
In this section, we describe the experimental setup for our study.

\begin{table}[t]
  \centering
  \caption{Summary of AI-authored commits by coding tool.}
    \scalebox{0.95}{
    \begin{tabular}{lrrr}
    \toprule
    AI Coding Tool\textsuperscript{\dag} & \# AI Commits & \# Repos & Avg. Commits/Repo \\
    \midrule
    GitHub Copilot & 118,012 & 4,487 & 26.3 \\
    Claude & 138,249 & 2,434 & 56.8 \\
    Cursor & 19,587 & 714 & 27.4 \\
    Gemini & 12,429 & 681 & 18.3 \\
    Devin & 14,302 & 298 & 48.0 \\
    \midrule
    Total & 302,579 & 6,299\textsuperscript{*} & 48.0 \\
    \bottomrule
    \end{tabular}%
    }
    \smallskip \\
    \footnotesize\textsuperscript{\dag}Each name groups all variants of the tool (e.g., \textit{Claude} includes Claude Code and its different model versions). \footnotesize\textsuperscript{*}Repositories may use more than one tool; the total is deduplicated.
  \label{tab:dataset_tool_summary}%
\end{table}%

\subsection{Dataset Summary}
\begin{figure}[t]
    \centering
    \begin{subfigure}[t]{0.45\linewidth}
        \centering
        \includegraphics[width=\linewidth]{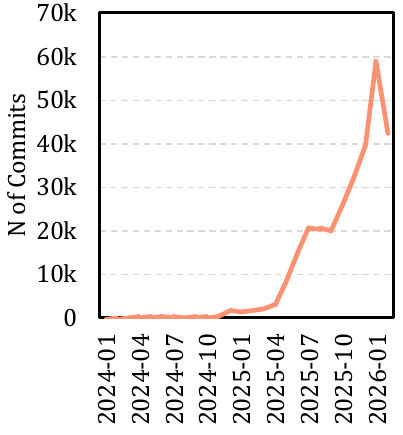}
        \caption{Monthly commits.}
        \label{fig:monthly_commits}
    \end{subfigure}
    \hfill
    \begin{subfigure}[t]{0.4\linewidth}
        \centering
        \includegraphics[width=\linewidth]{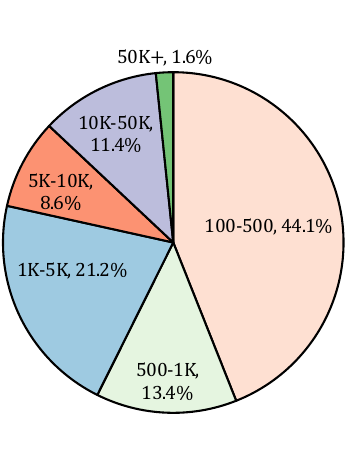}
        \caption{Repository distribution.}
        \label{fig:repo_stars}
    \end{subfigure}
    \caption{Overview of our dataset: (a) growth of AI-authored commits over time, and (b) distribution of repositories by GitHub star count as of March 2026.}
    \label{fig:dataset_overview}
\end{figure}

We collected our dataset by mining GitHub and applying filtering criteria described in Section~\ref{sec:data_collection}.
After repository-level filtering, we obtained 6,699 repositories with confirmed AI-authored commits, covering 29 AI coding tools.
However, some tools have very few commits, which may not provide reliable data for comparison.
Thus, we focus on the five assistants with more than 10,000 attributed commits: GitHub Copilot~\cite{githubcopilot}, Claude~\cite{claudecode}, Cursor~\cite{cursor}, Gemini~\cite{geminicodeassist}, and Devin~\cite{devin}.
This results in 6,412 repositories with 317.4K AI-attributed commits.
For the commit-level quality analysis, we further exclude cases that cannot be analyzed reliably such as repositories that became unavailable, deletion-only commits, and commits that do not modify production Python, JavaScript, or TypeScript source files.
Our final analysis dataset therefore includes \textbf{6,299} public GitHub repositories with \textbf{302.6K} analyzed AI-attributed commits.
Table~\ref{tab:dataset_tool_summary} summarizes the distribution of commits across the five AI coding assistants in our dataset.
Figure~\ref{fig:monthly_commits} shows the monthly growth of AI-authored commits, with a sharp increase starting from mid-2025.
Figure~\ref{fig:repo_stars} shows that our dataset covers repositories with a wide range of popularity levels.

\subsection{Research Questions}
\label{sec:research_questions}

Our study focuses on the following research questions:

\noindent\textbf{RQ1: What kinds of technical debt are introduced by AI coding assistants?}
This question investigates the basic characteristics of AI-introduced technical debt.
We study what types of debt they introduce, including code smells, correctness issues, and security issues.
We also analyze how these issues are distributed across languages, rules, and AI coding assistants.

\noindent\textbf{RQ2: How does technical debt vary across AI coding assistants?}
This question compares different AI coding assistants at the commit level.
We examine whether some assistants introduce more technical debt than others, and whether the kinds of issues they introduce differ.
This helps us understand whether technical debt patterns are tool-specific.

\noindent\textbf{RQ3: To what extent does AI-introduced technical debt persist in the codebase?}
Introducing debt is not necessarily a problem if it gets fixed quickly.
The real concern is debt that persists unnoticed.
We compare the number of issues introduced and fixed by AI-authored commits to assess the net impact.
We study whether introduced issues remain in the latest version of the repository or disappear over time.

\subsection{Evaluation Metrics}
\label{sec:eval_metrics}
We use the following metrics to support our three research questions.
For issue introduction (RQ1, RQ2), we report the total number of issues introduced by AI-authored commits, the percentage of commits that introduce at least one issue, and the average number of issues per commit.
We break these down by issue type, programming language, rule, and AI coding assistant.

For the debt lifecycle (RQ3), we use two metrics.
First, we compute the net impact by comparing the number of issues introduced and fixed by AI-authored commits.
Second, we measure the survival rate of introduced issues:
\[
\textit{Survival Rate} =
\frac{\text{\# issues surviving at \texttt{HEAD}}}
{\text{\# issues tracked}}
\]

\subsection{Validation}
\label{sec:validation}

To assess the reliability of our pipeline, two authors independently inspected random samples of AI-attributed commits and detected issues.

We randomly sampled 100 AI-attributed commits.
For each commit, we manually verified whether it was correctly attributed to the claimed AI coding assistant using the Git metadata and commit message described in Section~\ref{sec:data_collection}.
One sampled case could not be verified because the corresponding repository was no longer available.
For the remaining 99 verifiable commits, both authors independently labeled all cases, and all 99 were confirmed as correctly attributed after manual inspection.
This yields a conservative attribution precision of 99.0\% over the full sample, or 100\% over the verifiable subset.

We also randomly sampled 100 introduced issues from manually verified AI-attributed commits and validated two aspects of each:
(1) whether the reported issue was a real issue, rather than a false positive from the static analysis tools; and
(2) whether the survival classification at \texttt{HEAD} was correct (i.e., whether the issue was correctly labeled as surviving or not surviving).
After excluding one case with unavailable or incomplete validation context, both authors independently labeled 99 issue cases.
For issue validity, the raw agreement was 95/99 = 0.960, with Cohen's $\kappa = 0.851$, indicating almost perfect agreement.
For survival classification, the raw agreement was 97/99 = 0.980, with Cohen's $\kappa = 0.960$, also indicating almost perfect agreement.

Using the adjudicated labels as reference, the pipeline achieved an accuracy of 85.9\%, precision of 85.9\%, recall of 100.0\%, and F1 of 92.4\% for issue validity.
For survival classification, the pipeline achieved an accuracy of 84.8\%, precision of 86.7\%, recall of 81.2\%, and F1 of 83.9\%.
These results suggest that our pipeline provides reliable large-scale estimates, while still leaving some room for noise in issue detection and lifecycle tracking.
\section{Results}
\label{sec:results} 

This section presents the empirical results of our study and answers the three research questions raised in Section~\ref{sec:research_questions}.

\begin{table}[t]
  \centering
  \caption{Overview of AI-introduced technical debt by issue type.}
  \scalebox{0.9}{
    \begin{tabular}{lrrrr}
    \toprule
    \textbf{Type} & \textbf{\# Introduced} & \textbf{\# Repos} & \textbf{\# Commits} & \textbf{\% of Issues} \\
    \midrule
    Code Smells     & 432,748 & 3,812 & 25,740 & 89.3\% \\
    Correctness Issues    & 28,931 & 665   & 1,650 & 6.0\% \\
    Security Issues & 22,687 & 1,643 & 5,142 & 4.7\% \\
    \midrule
    \textbf{All Issues} & \textbf{484,366} & \textbf{3,946} & \textbf{27,677} & \textbf{-} \\
    \bottomrule
    \end{tabular}
  }
  \label{tab:rq1_issue_family}
\end{table}

\subsection{RQ1: Types and Patterns of AI-Introduced Debt}
\label{subsec:rq1}

\smallsection{Overview}
Table~\ref{tab:rq1_issue_family} presents a summary of the technical debt introduced by AI coding assistants in our dataset.
In total, we identified 484,366 introduced issues across 3,946 repositories (62.6\% of 6,299 repositories) and 27,677 commits (9.1\% of 302,579 commits).
This shows that a non-trivial portion of AI-authored commits introduce quality issues, and that these issues affect a large number of real-world repositories.
Among all introduced issues, code smells, correctness issues, and security issues are the three main categories.
Code smells are by far the most common, accounting for 89.3\% of all introduced issues.
Below, we discuss each type in detail with real-world examples.

\begin{table}[t]
  \centering
    \caption{Top 5 most frequent rules violated by AI coding assistants for each issue type.}
  \scalebox{0.92}{
    \begin{tabular}{llrr}
    \toprule
    \textbf{Type} & \textbf{Rule} & \textbf{Count} & \textbf{Rate} \\
    \midrule
    \multicolumn{1}{l}{\multirow{5}[2]{*}{Code Smells}} & Broad exception handling & 41,374 & 8.5\% \\
          & Unused variables or parameters & 28,272 & 5.8\% \\
          & Unused argument & 24,357 & 5.0\% \\
          & Shadowed outer variable & 20,647 & 4.3\% \\
          & Access to protected member & 19,796 & 4.1\% \\
    \midrule
    \multicolumn{1}{l}{\multirow{5}[2]{*}{Correctness Issues}} & Undefined variable or reference & 23,856 & 4.9\% \\
          & Redeclared symbol & 1,888 & 0.4\% \\
          & Possibly used before assignment & 1575  & 0.3\% \\
          & Access member before definition & 893   & 0.2\% \\
          & Unsubscriptable object & 176   & 0.04\% \\
    \midrule
    \multicolumn{1}{l}{\multirow{5}[2]{*}{Security Issues}} & Path traversal via path.join/resolve & 8,677 & 1.8\% \\
          & Unsafe format string & 4,792 & 1.0\% \\
          & Non-literal regular expression & 1,212 & 0.3\% \\
          & Child process execution & 607   & 0.1\% \\
          & SQLAlchemy raw query execution & 591   & 0.1\% \\
    \bottomrule
    \end{tabular}
  }
  \label{tab:rq1_top_rules}
\end{table}

\begin{listing}[t]
\centering

\begin{issuecode}{\texttt{ArchiveBox~\cite{archiveboxcommit_issue}}, \texttt{core/models.py:L594:598}, \texttt{d360798}}
# issue: broad exception handler silently swallows JSON-loading failures
try:
    with |\bugline{open(json\_path)}| as f:
        data = json.load(f)
|\bugline{except:}|
    pass
\end{issuecode}

\caption{Code smell: broad exception handling and missing file encoding in \texttt{ArchiveBox}~\cite{archiveboxrepo} (Claude Code).}
\label{lst:archivebox_example}
\end{listing}

\smallsection{Code Smells}
Code smells are maintainability problems that make code harder to understand, debug, and evolve~\cite{fowler2018refactoring}.
They increase long-term maintenance costs, even if they do not cause immediate failures.
This finding is consistent with prior work under controlled settings~\cite{liu2024refining, siddiq2024quality}, but our study confirms that the same pattern also appears in real-world repositories.
Table~\ref{tab:rq1_top_rules} lists the top 5 most common code smell patterns (e.g., broad exception handling, unused variables or parameters).
These issues are often small and easy to overlook during code review.
Listing~\ref{lst:archivebox_example} shows an example from \texttt{ArchiveBox}~\cite{archiveboxrepo} ($>$27k stars).
In commit \texttt{d36079829bed}~\cite{archiveboxcommit_issue}, Claude Code updated the metadata loading logic in \texttt{ArchiveBox}.
But the new code introduces two code smells.
First, the bare \texttt{except: pass} block catches all exceptions silently.
This makes errors harder to detect and debug.
Second, the \texttt{open()} function does not specify a file encoding.
This can lead to inconsistent behavior across different platforms and locales, since the default encoding may vary~\cite{pep597}.
These issues may not cause immediate failures, but they can lead to maintenance challenges and subtle bugs in the future.

\begin{listing}[t]
\centering

\begin{issuecode}{\texttt{firecrawl~\cite{firecrawlcommit_nameerror}}, \texttt{firecrawl.py:L4004:4047}, \texttt{fb99747}}
# issue: undefined variable "cache" causes a runtime NameError
async def generate_llms_text(
        self,
        url: str,
        *,
        max_urls: Optional[int] = None,
        show_full_text: Optional[bool] = None,
        experimental_stream: Optional[bool] = None) -> GenerateLLMsTextStatusResponse:

    response = await self.async_generate_llms_text(
        url,
        max_urls=max_urls,
        show_full_text=show_full_text,
        |\bugline{cache=cache,}|
        experimental_stream=experimental_stream
    )
\end{issuecode}

\caption{Correctness issues: undefined variable causing \texttt{NameError} in \texttt{Firecrawl}~\cite{firecrawlrepo} (Devin).}
\label{lst:firecrawl_nameerror_example}
\end{listing}

\smallsection{Correctness Issues}
Correctness issues are code defects that can cause the program to fail during execution.
Compared with code smells, they are less frequent.
From Table~\ref{tab:rq1_issue_family}, 28,931 correctness issues are identified, which cover 665 repositories and 1,650 commits.
However, their impact is more direct and severe than code smells.
Table~\ref{tab:rq1_top_rules} shows the top 5 most common correctness issues, which include undefined variable or reference, redeclared symbol, access to member before definition, possibly used before assignment, and unsubscriptable object.
These patterns suggest that AI-generated code may look locally correct, but still fail to stay consistent with the surrounding context.
What is interesting in this table is that we identified 23,856 cases of undefined variable or reference.
Listing~\ref{lst:firecrawl_nameerror_example} presents one such case from \texttt{firecrawl} ($>$98k stars).
In commit \texttt{fb99747ba978}~\cite{firecrawlcommit_nameerror}, Devin added a call that passes \texttt{cache=cache} as an argument.
However, \texttt{cache} is never defined in the method, which leads to a \texttt{NameError} when that path is executed.
The maintainer later fixed the bug by removing the undefined argument~\cite{firecrawlcommit_fix}.
This example shows that AI-generated code can introduce real correctness issues.
These errors require additional human effort to fix later.

\begin{listing}[t]
\centering

\begin{issuecode}{\texttt{data-formulator~\cite{dataformulatorcommit}}, \texttt{tables\_routes.py:L881:889}, \texttt{d8549c0}}
# issue: user-controlled table name is interpolated directly into SQL
for source_name in source_table_names:
    if source_name == updated_table_name:
        df = pd.DataFrame(updated_table['rows'])
    else:
        with db_manager.connection(session['session_id']) as db:
            |\bugline{\detokenize{result = db.execute(}}|
              |\bugline{\detokenize{f"SELECT * FROM {source_name}").fetchdf()}}|
            df = result
\end{issuecode}

\caption{Security issue: possible SQL injection in \texttt{microsoft/data-formulator}~\cite{dataformulatorrepo} (Copilot).}
\label{lst:dataformulator_sqli_example}
\end{listing}

\smallsection{Security Issues}
Security issues are another concern in AI-generated code.
In our study, this category includes not only direct security vulnerabilities, but also insecure coding patterns that can be viewed as security debt.
Some of these issues may be exploitable at the time they are introduced, while others may become security risks after later code changes or broader system integration.
Thus, it is important to identify and fix these issues early before they accumulate in production repositories.
As shown in Table~\ref{tab:rq1_issue_family}, potentially insecure code patterns are detected in 1,643 repositories and 5,142 commits.
Table~\ref{tab:rq1_top_rules} shows that common security issues such as path traversal via \texttt{path.join} or \texttt{path.resolve}, unsafe format strings, non-literal regular expressions, and child process execution.
These patterns suggest that AI-generated code can introduce unsafe practices in process execution, file path handling, and string formatting.
A common pattern across these issues is unsafe handling of untrusted input, where user- or context-controlled values flow into security-sensitive operations without proper validation or sanitization.
Figure~\ref{fig:hysteria2_example} shows one such example from \texttt{hysteria2} ($>$1.5k stars).
A Copilot-authored commit~\cite{hysteria2commit_issue} introduced a \texttt{shell=True} subprocess call.
This pattern is not necessarily an exploitable vulnerability by itself in the current version, but it creates security debt because it can enable command injection if untrusted input later reaches the command string.
A human developer later identified and removed the unsafe flag~\cite{hysteria2commit_fix}.
Beyond that, Listing~\ref{lst:dataformulator_sqli_example} shows another example of a security issue (SQL injection) in \texttt{data-formulator} ($>$1.2k stars).
This repository is developed by Microsoft and uses GitHub Copilot for code generation.
In commit \texttt{d8549c0}~\cite{dataformulatorcommit}, Copilot added a backend endpoint that constructs a SQL query by directly interpolating a user-supplied table name.
This creates a potential SQL injection vector.
If an attacker can control the \texttt{source\_name} variable, they can inject malicious SQL code that may lead to data breaches or unauthorized access.
This issue remained in the repository for several weeks before the maintainer refactored the code and removed the unsafe SQL construction~\cite{dataformulatorrepo}.

\begin{table}[t]
  \centering
  \caption{Top 5 most frequent rules by language. Rate indicates each rule's share of all issues in that language.}
  \scalebox{0.95}{
    \begin{tabular}{llrr}
    \toprule
    \textbf{Language} & \textbf{Rule} & \textbf{Count} & \textbf{Rate} \\
    \midrule
    \multirow{5}{*}{Python}
          & Broad exception handling & 41,374 & 14.9\% \\
          & Unused argument & 24,357 & 8.8\% \\
          & Undefined variable or reference & 23,856 & 8.6\% \\
          & Access to protected member & 19,796 & 7.1\% \\
          & Unused import & 17,376 & 6.3\% \\
    \midrule
    \multirow{5}{*}{\shortstack[l]{JavaScript/\\TypeScript}}
          & Unused variables or parameters & 28,272 & 13.6\% \\
          & Shadowed outer variable & 18,417 & 8.9\% \\
          & Block-scoped variable misuse & 11,568 & 5.6\% \\
          & No sequences & 9,358  & 4.5\% \\
          & Path traversal via path.join/resolve & 8,677  & 4.2\% \\
    \bottomrule
    \end{tabular}
  }
  \label{tab:rq1_top_rules_lang}
\end{table}

\smallsection{Programming Language Differences}
Table~\ref{tab:rq1_top_rules_lang} compares the top 5 most frequent rules in Python and JavaScript/TypeScript.
There are some patterns that are common in both languages.
For example, both languages show issues related to unused code (i.e., unused arguments in Python, unused variables in JavaScript/TypeScript).
At the same time, each language also has its own characteristic issues.
Python's top rules are dominated by exception handling and dynamic typing problems, while JavaScript/TypeScript issues tend to involve scoping and variable declaration patterns.
This observation is consistent with prior studies~\cite{liu2024refining, siddiq2024quality}, which also found that the types of issues in AI-generated code can vary depending on the programming language and the tools used.
These results suggest that some debt patterns may be language-specific.
However, the overall trend (e.g., code smells dominate) holds across both languages.

\find{
\textbf{Answer to RQ1:}
AI-generated code introduces technical debt in the form of code smells (89.3\%), correctness issues (6.0\%), and security issues (4.7\%).
Among them, code smells are by far the most common.
}

\begin{figure}
    \centering
    \includegraphics[width=0.7\linewidth]{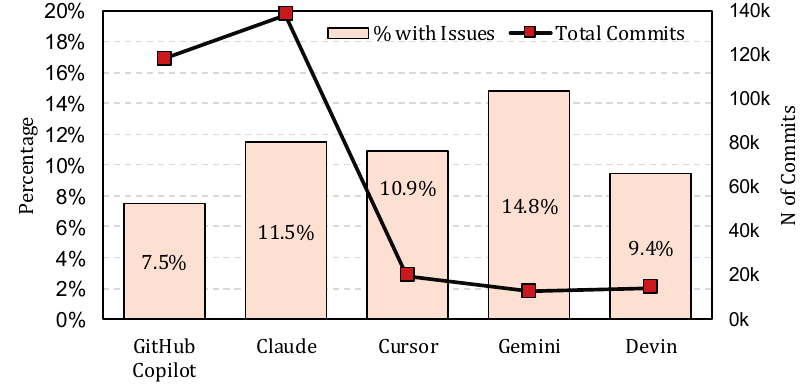}
    \caption{Percentage of commits with issues and total commit volume per AI coding assistant.}
    \label{fig:rq2_coding_compare}
\end{figure}

\begin{table}[t]
    \centering
    \includegraphics[width=0.75\linewidth]{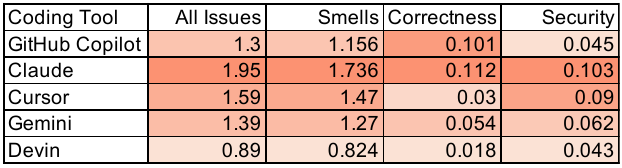}
    \caption{Average number of issues introduced per commit by type. Darker cells indicate higher values.}
    \label{tab:rq2_coding_compare_per_c}
\end{table}

\subsection{RQ2: Comparison Across AI Coding Assistants}
In this RQ, we examine how technical debt patterns vary across AI coding assistants.
Figure~\ref{fig:rq2_coding_compare} shows the percentage of commits with issues for each of the five AI coding assistants.
Across the five AI coding assistants, the rates range from 7.5\% for GitHub Copilot to 14.8\% for Gemini.
Claude, Cursor, and Devin have corresponding rates of 11.5\%, 10.9\%, and 9.4\%, respectively.
This suggests that technical debt appears across all studied tools, although the rate differs by tool.
In other words, the problem is not limited to a single AI coding assistant.

Table~\ref{tab:rq2_coding_compare_per_c} further compares the average number of introduced issues per commit by type.
We can see that all five tools share a common pattern, where the code smell rate is much higher than the correctness and security issue rates.
This is consistent with our findings in RQ1.
At the same time, there are also differences across tools.
From Table~\ref{tab:rq2_coding_compare_per_c}, we can see that Claude has the highest issue rate per commit (1.95), while Devin has the lowest (0.89).
These differences may be due to differences in usage patterns and development context, rather than the tools alone.
Still, it is apparent that the overall pattern of technical debt is consistent across all five tools.

\find{
\textbf{Answer to RQ2:}
Technical debt patterns vary across AI coding assistants, but the overall trend is consistent.
Across all five tools, code smells remain the dominant form of AI-introduced debt.
}

\begin{figure}
    \centering
    \includegraphics[width=0.8\linewidth]{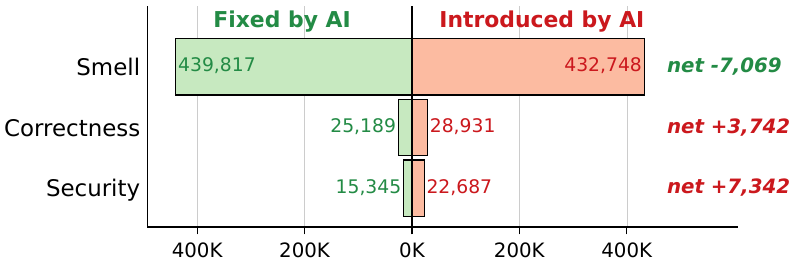}
    \caption{Net impact of AI coding assistants: issues introduced vs. fixed by issue type.}
    \label{fig:rq3_butterfly_net}
\end{figure}

\subsection{RQ3: Persistence of AI-Introduced Debt}
\label{sec:rq3}

\smallsection{Net Impact}
In RQ1 and RQ2, we focus on the technical debt introduced by AI-authored commits.
However, AI coding assistants can also remove existing issues during refactoring or code improvement.
To better understand the overall lifecycle of AI-introduced debt, we compare the number of issues introduced and fixed by AI commits (see Figure~\ref{fig:rq3_butterfly_net}).
For code smells, we can see that AI-authored commits fix more issues than they introduce (439,817 vs. 432,748), resulting in a net reduction of 7,069 code smells.
In contrast, for correctness and security issues, AI commits introduce more issues than they fix.
What is interesting is that AI introduces about 1.5 times as many security issues as it fixes.
These findings indicate that the net impact of AI coding assistants is mixed.
AI coding assistants can help reduce maintainability issues, which tend to follow simple and repetitive patterns.
However, for correctness and security issues, which require a deeper understanding and reasoning about program logic and context, AI coding assistants introduce more problems than they resolve.

\begin{figure}
    \centering
    \includegraphics[width=0.65\linewidth]{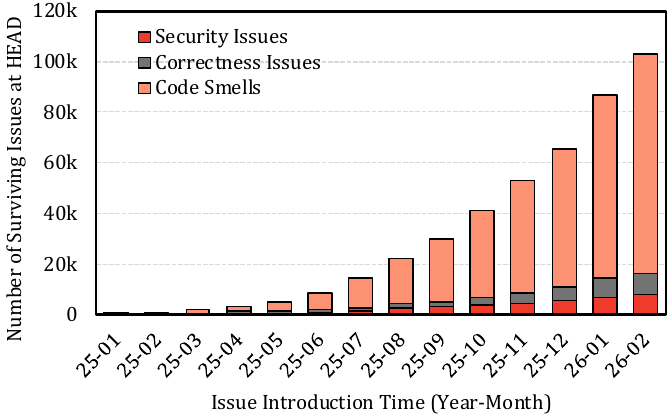}
    \caption{Cumulative growth of AI-introduced issues over time, by issue type.}
    \label{fig:rq3_cumulative_issue}
\end{figure}

\begin{table}[t]
  \centering
  \caption{Survival of AI-introduced issues by time since introduction.}
  \scalebox{0.9}{
    \begin{tabular}{lrrrr}
    \toprule
    \textbf{Time Since Introduction} & \textbf{\# Commits} & \textbf{\# Issues} & \textbf{\# Surviving} & \textbf{Survival Rate} \\
    \midrule
    $>$ 9 months & 17,612 & 21,421 & 4,893 & 22.8\% \\
    6--9 months & 56,955 & 89,135 & 17,307 & 19.4\% \\
    3--6 months & 79,986 & 111,551 & 31,492 & 28.2\% \\
    $<$ 3 months & 147,979 & 242,793 & 51,672 & 21.3\% \\
    \midrule
    \textbf{All Issues} & \textbf{302,532} & \textbf{464,900} & \textbf{105,364} & \textbf{22.7\%} \\
    \bottomrule
    \end{tabular}
  }
  \label{tab:rq3_survival_age}
\end{table}

\smallsection{Issue Survival}
The net impact analysis above provides an overview of what AI coding assistants add and remove.
But it does not show what happens to the specific issues introduced by AI.
To answer this question, we track each AI-introduced issue to the latest repository snapshot and check whether it still exists at \texttt{HEAD}.
Figure~\ref{fig:rq3_cumulative_issue} shows that the cumulative number of surviving issues keeps growing over time.
The total volume of unresolved technical debt increases rapidly, climbing from just a few hundred issues in early 2025 to over 100k surviving issues by February 2026.
This suggests that as the rapid adoption of AI coding assistants continues, the amount of AI-introduced debt in real-world repositories is also growing significantly.

Table~\ref{tab:rq3_survival_age} provides an age-cohort view of issue survival.
Overall, 105,364 out of 464,900 tracked AI-introduced issues~\footnote{Table~\ref{tab:rq3_survival_age} reports available survival-analysis results; survival data were unavailable for 23 repositories.} still survive at \texttt{HEAD}, corresponding to a survival rate of 22.7\%.
Surviving issues appear in all age cohorts, including issues introduced more than nine months earlier.
For example, 4,893 issues introduced more than nine months ago still remain at \texttt{HEAD}.
The survival rate varies across cohorts, ranging from 19.4\% for issues introduced 6--9 months ago to 28.2\% for issues introduced 3--6 months ago.
This suggests that AI-introduced debt is not always removed quickly after it enters the codebase.
Although the cohort-level survival rates do not show a simple monotonic trend, the main finding is clear: a substantial number of AI-introduced issues remain unresolved over time.

\begin{figure}[t]
    \centering    
    \includegraphics[width=0.7\textwidth]{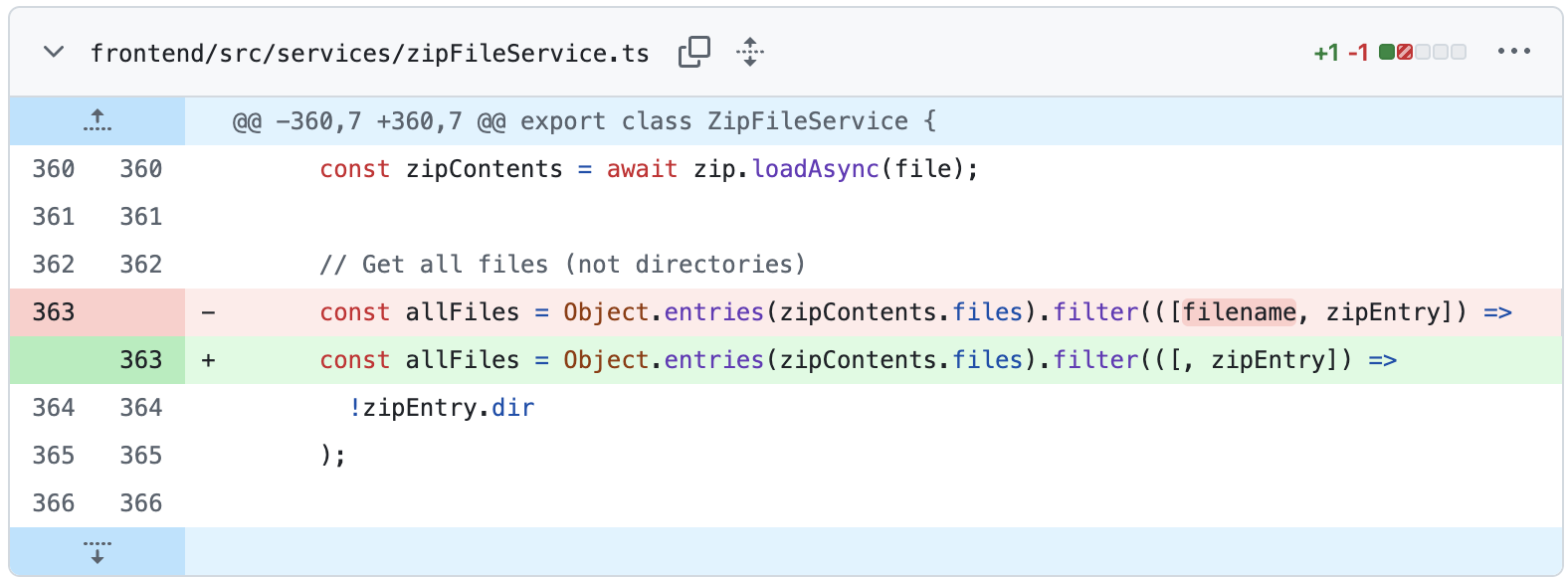}
    \caption{A TypeScript lint issue introduced by a Claude-authored commit in \texttt{Stirling-PDF} was fixed one day later by removing the unused variable.}
    \label{fig:typescipt_fix_example}
\end{figure}

These aggregate results are also reflected in real-world repositories.
For example, the broad exception handling issue in Listing~\ref{lst:archivebox_example} was fixed within hours after it was introduced~\cite{archiveboxcommit_issue,archiveboxcommit_fix}.
Similarly, in \texttt{Stirling-PDF}~\cite{stirlingpdfrepo} ($>$75k stars), a Claude-authored commit introduced an unused variable \texttt{filename} in a TypeScript file~\cite{stirlingpdfcommit_issue}.
As shown in Figure~\ref{fig:typescipt_fix_example}, the maintainer fixed it the next day with a commit titled ``Fix TypeScript linting error''~\cite{stirlingpdfcommit_fix}.
In contrast, the undefined variable bug in \texttt{firecrawl} (Listing~\ref{lst:firecrawl_nameerror_example}) took 42 days before the maintainer fixed it~\cite{firecrawlcommit_fix}.
In some cases, some issues can survive for much longer or even remain unresolved.
For instance, a Devin-authored commit in \texttt{brave\_search\_tool.py} added a call to \texttt{requests.get(...)} without a timeout in December 2024~\cite{crewai439cde}.
This is a known potential security issue, because requests made without a timeout may block indefinitely if the remote service does not respond~\cite{bandit_b113}.
However, the issue still remains in the latest repository revision.

\find{
\textbf{Answer to RQ3:}
AI-authored commits fix slightly more code smells than they introduce, but introduce more correctness and security issues than they fix.
Overall, 22.7\% of tracked AI-introduced issues still survive at \texttt{HEAD}, including issues introduced more than nine months earlier.
}

\section{Discussion}
\label{sec:discussion}
 
\subsection{Implications}
Our empirical study shows that AI coding assistants introduce technical debt into real software repositories.
This is not a property of any single tool.
We observed that across all five tools we studied, around 9\% of commits introduce at least one detectable issue.
These issues persist regardless of repository size or popularity.
In this section, we discuss the implications of our findings for practitioners, researchers, and tool builders.

\textbf{AI-assisted development creates persistent debt, not just temporary low-quality code.}
The main implication is not just that AI coding assistants may produce low-quality code.
More importantly, AI-assisted software development changes how technical debt enters and remains in production systems.
Prior studies have shown that developers are more likely to over-trust AI suggestions~\cite{sabouri2025trust}.
This over-trust can lead to a higher acceptance rate of AI-generated code, even when it contains issues.
It means that many AI-introduced issues can accumulate in the codebase.
Our results show that code smells are the most common type of AI-introduced debt.
They often do not break the software system immediately, making them easy to accept during code review.
But AI coding assistants allow developers to produce code at a much higher speed and volume.
As a result, these minor issues can accumulate into a substantial maintenance burden over time.
Figure~\ref{fig:rq3_cumulative_issue} shows that the cumulative number of surviving AI-introduced issues continues to rise over time, exceeding 100k by February 2026.
The technical debt introduced by AI does not have to be a temporary side effect, but a long-term maintenance challenge for modern software systems.

\textbf{Developers should be especially cautious about correctness and security issues.}
Our findings suggest that although AI coding assistants introduce technical debt, they also fix existing issues in the codebase.
First, we observe that AI co-authored commits actually fix a similar number of code smell issues as they introduce.
This suggests that AI coding assistants are able to perform local cleanup and repetitive maintenance tasks effectively. 
They can recognize and address surface-level code quality problems (e.g., formatting, naming, or simple refactoring opportunities).
However, what is concerning is that AI coding assistants seem to be less effective at fixing correctness and security issues, and they even introduce more of these than they fix.
Section~\ref{sec:rq3} shows that many AI-introduced issues still survive at \texttt{HEAD}, while the net-impact analysis shows that AI commits introduce more correctness and security issues than they fix.
This inconsistency suggests that AI coding assistants may struggle with changes that require a deeper understanding of program behavior, execution context, or security implications.
Also, the practical impact of correctness and security issues is often more severe than that of code smells, making it more critical to address them effectively.
Developers should not treat all AI-generated code as equally trustworthy, and they should pay particular attention to changes that may introduce correctness or security vulnerabilities.

\textbf{Technical debt cannot be solved by switching between AI coding tools.}
Our cross-tool comparison shows that this problem cannot be solved simply by switching from one assistant to another. 
Figure~\ref{fig:rq2_coding_compare} shows that all five tools introduce a similar pattern of issues.
They all have a high rate of code smells, and a non-trivial rate of correctness and security issues.
This suggests that the quality risk is a systemic issue with the current mode of AI-assisted development.
Thus, developers and teams should carefully review AI-generated code regardless of the tool used.
Static analysis, tests, and security checks should be part of the normal workflow.
Code review should extend beyond the point of merge.
The main reason is that our study shows that 22.7\% of tracked AI-introduced issues still survive at \texttt{HEAD}, and older issues are not fully cleaned up even after months.
Merging the AI-generated code does not mean the end of the story since debt can persist and accumulate over time.
This makes continuous monitoring and targeted debt repayment necessary for AI-touched code.

\textbf{Future research and tool design should target long-term code health.}
Our results suggest that future research and tool design should not just focus on generating more acceptable code.
We also need to ask whether that code remains maintainable, correct, and secure over time. 
Most existing research studies~\cite{liu2024refining, siddiq2024quality} focus on short-term outcomes such as task completion, acceptance rate, or immediate correctness.
However, these measures only capture what happens when the code is introduced, not what happens later in maintenance.
Future work needs to examine which factors make AI-introduced debt more likely to persist (e.g., repository maturity, review intensity, or task type).
Our findings also suggest the need for better assistants. Future tools should make stronger checks for security-sensitive changes, use repository context more effectively, and clearly show AI provenance so reviewers can better judge the risk of a change. In the end, the key question is not only whether AI can produce code at scale, but whether the software engineering ecosystem can manage that code well over time.

\section{Related work}
\label{sec:relatedwork}

\smallsection{Code Quality of AI-Generated Code}
Previous studies have examined the quality and security of AI-generated code.
Based on controlled experiments, they showed that AI coding assistants (e.g., GitHub Copilot and ChatGPT) are able to produce functional code, but the code quality varies widely across languages, tasks, and prompts~\cite{nguyen2022empirical,liu2024refining, mastropaolo2023robustness}. 
AI-generated code can also contain security weaknesses, and developers may over-trust it and fail to properly review it~\cite{pearce2025asleep, perry2023users, sandoval2023lost}.
Recent studies have also begun examining AI-generated code in real-world production environments.
They show that AI-generated code is being widely adopted in platforms such as GitHub and Stack Overflow, and that it can carry quality and security issues~\cite{fu2025security, siddiq2024quality, wang2025ai}.
He~\ea~\cite{he2025does} studied the impact of Cursor adoption on 807 repositories and observed persistent increases in code complexity.
Watanabe~\ea~\cite{watanabe2025use} found that most Claude Code pull requests are merged, though many require human revisions.

Together, these studies show that AI-generated code can contain quality and security problems in both controlled and real-world settings.
However, these studies mainly focus on a single tool or a narrow set of quality issues, and they do not track how those issues evolve.
In contrast, our work provides a comprehensive analysis of the quality and security of AI-generated code across multiple tools, and tracks how those issues persist in production repositories over time.

\smallsection{Empirical Studies of AI-Assisted Development Practices}
Recent studies have examined how AI coding assistants are used in real software development practice.
Several studies examined the productivity impact of these tools.
Peng~\ea~\cite{peng2023impact} conducted a randomized controlled trial and found that developers using GitHub Copilot completed tasks 55.8\% faster.
At industrial scale, Ziegler~\ea~\cite{ziegler2024measuring} and Murali~\ea~\cite{murali2024ai} reported that suggestion acceptance rate strongly correlates with self-reported productivity, with acceptance rates around 22--30\%.
Other studies focus on usage patterns and developer perceptions.
Liang~\ea~\cite{liang2024large} surveyed 410 developers and found that developers mainly use AI assistants to reduce keystrokes and recall syntax, but often reject suggestions that fail to meet functional requirements~\cite{davila2024industry}.
Sabouri~\ea~\cite{sabouri2025trust} further observed that developers keep only 52\% of AI suggestions after review, and Klemmer~\ea~\cite{klemmer2024using} found that developers widely use AI assistants for security-critical tasks despite concerns about suggestion quality.

These studies improve our understanding of how developers use AI assistants in practice.
However, they mainly focus on adoption, usability, trust, productivity, and collaboration, rather than on the technical debt carried by AI-authored code.
In contrast, our work examines the code-level debt introduced by AI coding assistants and studies whether that debt persists in production repositories over time.

\smallsection{Technical Debt in Software Development}
Technical debt has been a core topic in software engineering research for years.
Previous work has introduced many automated methods to identify code smells and self-admitted technical debt in software repositories~\cite{yan2018automating, ren2019neural, palomba2018diffuseness}.
Beyond detection, many studies have also investigated the lifecycle of technical debt in human-written code.
Tufano~\ea~\cite{tufano2017and} showed that code smells are often introduced during normal development. 
Once introduced, they can linger in the codebase for a long time before anyone removes them~\cite{tufano2017and}.
Digkas~\ea~\cite{digkas2018developers} identified a similar pattern in large open-source ecosystems.
Their results showed that technical debt mainly accumulates when new code is added~\cite{digkas2018developers, digkas2020can}.
Other studies further suggest that debt is rarely removed in an intentional way, and that automated repayment is still difficult in practice~\cite{zampetti2018self, mastropaolo2023towards}.

Prior studies provide a strong foundation for understanding technical debt in human-written software.
However, it remains unclear whether AI-generated code follows the same patterns, and whether the same tools and techniques can be applied to it.
Our work fills this gap by analyzing the technical debt in AI-generated code and exploring how it persists in production repositories over time.

\section{Threats to Validity}
\label{sec:threats}

Below, we discuss threats that may impact the results of our study.

\noindent\textbf{External Validity.}
Our study focuses on public GitHub repositories with at least 100 stars and production source files in Python, JavaScript, and TypeScript.
Therefore, our findings may not generalize to private repositories, smaller projects, or software written in other languages.
In addition, we only analyze AI-authored commits that leave explicit traces in Git metadata (e.g., bot actor logins, AI author emails, or \texttt{Co-authored-by} trailers).
AI-assisted contributions that leave no such trace are outside the scope of our dataset.
This design is analogous in spirit to research on self-admitted technical debt (SATD), which studies the subset of technical debt that developers explicitly document rather than the full universe of debt~\cite{potdar2014exploratory, da2017using}.
Like SATD research, our findings characterize a visible and attributable subset rather than the entire population of AI-assisted work.
We also do not compare AI-authored commits against a baseline of purely human-written commits.
Because developers may use AI without leaving any Git trace, and even AI-labeled commits may mix human edits, a reliable human-only baseline is difficult to construct, and comparing against an unreliable baseline could bias the results.
We therefore focus on tracking technical debt inside explicitly confirmed AI-authored or co-authored commits.

\noindent\textbf{Internal Validity.}
Our pipeline depends on the correctness of AI attribution, issue matching, and lifecycle tracking.
Although we use explicit Git metadata rather than proxy signals, some commits may still include both AI and human contributions.
Similarly, matching issues across revisions is challenging because files evolve over time and issues may shift location.
To reduce this risk, we compare code before and after each commit, restrict introduced issues to changed lines, and use rule identifiers, messages, and surrounding code context during matching.
For the issue survival analysis, a file may be deleted or entirely rewritten between the introducing commit and \texttt{HEAD}.
In such cases, the issue is classified as resolved, even though the resolution may not be a deliberate fix.
Our reported survival rates may therefore slightly underestimate the true persistence of AI-introduced debt.

\noindent\textbf{Construct Validity.}
Technical debt is a broad concept with many possible forms.
In this study, we operationalize technical debt mainly through code smells, correctness issues, and security issues detected by static analysis tools.
This choice allows us to measure debt consistently at scale and track it over time at the commit level.
However, static analysis tools can produce false positives, flagging code that is technically correct but matches a known risky pattern.
In addition, our security findings include both active vulnerabilities and latent unsafe patterns (i.e., security debt).
We do not separate these quantitatively, but Section~\ref{subsec:rq1} illustrates both cases.
Static analysis also does not cover all forms of technical debt.
Architectural debt, design erosion, documentation debt, and test adequacy issues are outside the scope of our tools.
Our results should therefore be interpreted as evidence about code-level technical debt, not the full spectrum of quality challenges that AI-generated code may introduce.

\section{Conclusion and Future Work}
\label{sec:conclusion}

AI coding assistants are rapidly becoming part of real-world software development, but their long-term impact on software quality remains unclear. 
In this paper, we presented a large-scale empirical study of the technical debt introduced by AI-generated code in the wild. 
By mining 302.6K AI-authored commits from 6,299 GitHub repositories, we designed a commit-level pipeline to identify introduced technical debt and track its later evolution in production repositories. 
Our study provides a longitudinal view of AI-generated code after it is merged. 
We look at what kinds of debt appear, how they differ across assistants, and whether they still remain at the latest revision. 
The results reveal the hidden maintenance costs behind AI coding assistants. They also show why stronger quality checks are needed in AI-assisted software development.
In future work, we plan to extend our analysis to other forms of debt (e.g., architectural, documentation, and test-related debt), additional programming languages, and broader development contexts.
We also plan to study what factors make AI-introduced debt more likely to persist.
Finally, we hope our findings can inform the design of more debt-aware AI coding assistants that help developers catch and fix quality issues before they enter production.


\bibliographystyle{ACM-Reference-Format}
\bibliography{sample-base}

\end{document}